\documentclass[a4paper,11pt]{article}
\usepackage{pos}
\usepackage{amsmath, amsthm, amsfonts}
\usepackage{hyperref}
\usepackage{slashed}
\usepackage{cleveref}
\usepackage{bbm}
\usepackage{braket}

\title{Spinning particles and background fields}

\author*[a,b]{Eugenia Boffo}

\affiliation[a]{Riemann Center for Geometry and Physics,\\
Leibniz Universit\"{a}t Hannover,\\
Appelstrasse 2, 30167 Hannover, Germany}
\affiliation[b]{Charles University Prague,\\
Faculty of Mathematics and Physics,  Mathematical Institute,\\
Sokolovsk\'{a} 49/83, 186 75 Prague, Czechia}

\emailAdd{boffo@karlin.mff.cuni.cz}

\abstract{Through their respective sigma models, a bosonic string and a superstring can be coupled to (super)gravity fields. These are subsequently forced to satisfy their right classical equation of motions, as a consequence of quantization of the string. There are indications that particle models with extended supersymmetry can replicate this behavior. The bosonic sector of supergravity, comprising the metric, the Kalb-Ramond 2-form and the dilaton scalar field, was already shown to derive from Becchi-Rouet-Stora-Tyutin quantization of the $N=4$ spinning particle \cite{Bonezzi:2020jjq}. Expanding on these results, here we discuss how to retrieve other Supergravity fields in the background.} 

\FullConference{%
  Corfu Summer Institute 2022 "School and Workshops on Elementary Particle Physics and Gravity",\\
  28 August - 1 October, 2022\\
  Corfu, Greece}

\tableofcontents

\begin{document}
\maketitle

\section{Introduction}

The purpose of this short note is to streamline some of the results on \emph{Ramond-Ramond backgrounds for the spinning particle} contained in \cite{Boffo:2022pbs}. These results were presented by the author in a talk at the Corfu Summer Institute, during the workshop "Noncommutative and generalized geometry in string theory, gauge theory and related physical models".~Emphasis is put on a representation with twistor-like variables and the need of a spin field for the study of deformations by backgrounds is justified. To begin with, I will first review $N=1$ spinning particles. 

\subsection{Spinning particle}

Spinning particles \cite{Brink:1976uf} are particles with an intrinsic spin degree of freedom. For our purposes we will focus on relativistic and massless spinning particles, but generalizations are possible (see for example \cite{Carosi:2021wbi} for the massive case). Supersymmetry hides behind the model and is responsible for the spin degree of freedom as we shall show. A reader familiar with string theory will recognize that the spinning particle is reminiscent of a superstring in the RNS formulation \cite{Neveu:1971iv} \cite{Ramond:1971gb}, when the 1-dimensional string collapses to a point-like object. To write down an action functional for the spinning particle we need to consider maps from the supermanifold $\mathbb{R}^{1\vert 1}$ (\emph{source space}) to a \emph{target space}   $(M,\eta)$, which we take to be a 4-dimensional metric manifold, Lorentzian for concreteness, with the Minkowski metric $\eta = \text{diag}(+--\,-)$. The superscript of $\mathbb{R}^{1\vert 1}$ counts and distinguishes the number of even and odd directions, where the latter are parametrized by Grassmann coordinates. So the function sheaf is locally isomorphic to $C^{\infty}(U) \otimes \Lambda^{\bullet} V$ (here $V$ is a 1-dimensional vector space and $U$ is an open subset of $\mathbb{R}$) or equivalently $C^\infty(\mathbb{R}) \oplus C^\infty(\mathbb{R}, V)$. The $\mu$-th component of an element in $\text{Maps}(\mathbb{R}^{1\vert1}, M)$ is thus:
\[
 X^\mu(\tau, \xi) = x^\mu(\tau) + \mathrm{i} \xi \psi^\mu(\tau)
\]
and plays the role of a coordinate for $M$, but it will not be the only datum entering the model. In fact, it is convenient to enforce \emph{reparametrization invariance} on the model. In the bosonic setting, reparametrization invariance is assured by an einbein $e(\tau)$. Hence we need a 1-form that generalizes $e(\tau)$ to $\mathbb{R}^{1\vert 1}$:
\[
 E(\tau, \xi) = e(\tau) + 2 \mathrm{i} \xi \chi(\tau) \in \Omega^1(\mathbb{R}) \oplus \Omega^1(\mathbb{R}, V).
\] 
Besides the natural vector fields $\partial_\tau $ and $\partial_\xi$, another useful derivation that we wish to consider here is the superderivative $D$:
\[
D = \mathrm{i} \xi \partial_\tau + \partial_\xi \, ,
\]
whose anticommutator closes on $\partial_\tau$. After all this preparation, we can assemble these ingredients into the action for the relativistic massless spinning particle, manifestly invariant under reparametrizations of the (super)line:
\begin{equation}
	S = - \int_{\mathbb{R}^{1 \vert 1}} \mathrm{d}\tau \mathrm{d}\xi \; \frac{\mathrm{i}}{2 E} D X^\mu \partial_\tau X^\mu
\end{equation} 
This functional integral bears evident similarities with its "bosonic" counterpart. Invariance under reparametrizations is simply the request that the Lie derivative of the Lagrangian w.r.t. a supervector field $Y(\tau, \xi) D$ is zero modulo boundary terms. On the fields, this corresponds to $ \delta_Y E = \mathcal{L}_{Y} E = Y (D E(\tau,\xi)) + (D Y(\tau, \xi)) E, \; \delta_Y X = Y (DX(\tau, \xi))$.

Expanding to linear order in $\xi$ and subsequently performing the Berezinian integral, one obtains:
\begin{equation}
	S = \int_{\mathbb{R}^{1}} \mathrm{d}\tau \; \frac{1}{2 e} \left(\dot{x}^\mu \dot{x}_\mu + \mathrm{i} \psi^\mu \dot{\psi}_\mu \right) - \frac{1}{e^2} \mathrm{i} \chi \psi^\mu \dot{x}_\mu
\end{equation}
This system is constrained. Through a Legendre transformation, the constraints can be conveniently emphasized. Thus with the Lagrangian $L$ in the \emph{first order formulation} ($L = p \cdot \dot{x} - H(p,x,\tau)$):
\begin{equation}
 L = p \cdot \dot{x} + \frac{\mathrm{i}}{2}\psi \cdot \dot{\psi} - \frac{e}{2} p^2 - \mathrm{i} \chi \psi \cdot p, \qquad \dot{x}^\mu = \left(e p_\nu + \frac{\mathrm{i}}{e} \chi \psi_\nu \right)\eta^{\mu\nu} \, ,  
\label{expr:L=px-H}
\end{equation} 
the constraints can be immediately read off: The Lagrange multiplier $\chi$ implements transversality of the bosonic momentum and the fermionic coordinate, while the einbein $e$ tells us that the Casimir element for translations must be zero, which is physically a zero mass condition. The infinitesimal supersymmetry transformations with local, parity odd parameter $\alpha(\tau)$:
\[
\delta_\alpha x^\mu = \mathrm{i} \, \alpha \psi^\mu \, , \quad \delta_\alpha \psi^\mu = - \alpha \eta^{\mu\nu} p_\nu \, , \quad \delta_\alpha e = 2 \mathrm{i} \, \alpha \chi \, , \quad  \delta_\alpha \chi = \dot\alpha \, , \quad \delta_\alpha p_\mu =0 \, ,
\]
leave the action functional invariant modulo a boundary term, $\delta_\alpha S = \int \mathrm{d} \tau \frac{\mathrm{i}}{2} \frac{\mathrm{d}}{\mathrm{d}\tau}(\alpha \psi^\mu p_\mu)$. From the Lagrangian in \eqref{expr:L=px-H}, the graded Poisson brackets are also immediate:
\[
\{x^\mu, p_\nu\} = \delta^\mu_\nu, \quad \{\psi^\mu, \psi^\nu\}_+ = 2\eta^{\mu\nu}.
\]
Given these brackets, evidently the constraints respect a superalgebra (associated to the superdiffeomorphism algebra)
\begin{equation}
H := p^2, \quad q := \psi \cdot p, \qquad \{q,q\} = 2H \, .
\label{symmetries}
\end{equation}

Canonical quantization of the system is straightforward at this stage. Turning the Poisson brackets into commutators of operators ($\hbar =1$), $p$ then acts by derivative action $p= - \mathrm{i} \frac{\partial}{\partial x}$, $x$ by multiplicative action and $\psi$ may have an action on two different modules at our disposal: the Spin module and the ring of forms. Let us explain how. The discussion that follows can also be found in \cite{Brink:1976uf}, one of the pioneering works on the subject.

\smallskip

\subsubsection{Spinors} 
Due to the Clifford algebra satisfied by the $\psi$'s, a natural module for the algebra is the module of spinors $S$. A 4-dimensional complex Dirac spinor $u(x) = (u_\alpha(x), u^{\dot\alpha}(x))^T$ is an irreducible representation of the Clifford algebra (though it is not irreducible for the Spin group, for which 2-dimensional Weyl spinors are). Then acting with the "transversality" constraint $q$ yields:
\[
 q u(x) = \begin{pmatrix}
 	\Gamma^{\mu}_{\alpha \dot\beta} \partial_\mu u^{\dot\beta}(x)\\
 	\Gamma^{\mu \, \dot\alpha \beta} \partial_\mu u_{\beta}(x) 
 \end{pmatrix} = \mathrm{i} \slashed{\partial} u  = 0 .
\]
This is the Dirac equation and $q$ was the Dirac operator all along. Furthermore the condition $p^2 = - \square =0 $ is automatically satisfied when the spinor already belongs to the kernel of the Dirac operator.

\smallskip

\subsubsection{Forms} 
\label{sec:Forms}

In the following part, our metric manifold $(M, \eta)$ will be locally identified with $\mathbb{R}^{1,3}$. The Clifford algebra $\text{Cliff}(\mathbb{R}^{1,3})$, with symmetric pairing of signature $(1,3)$, as a vector space is isomorphic to:
\[
\text{Cliff}(\mathbb{R}^{1,3}) \cong \Lambda^0 \mathbb{R}^{1,3} \oplus \Lambda^1 \mathbb{R}^{1,3} \oplus \Lambda^2 \mathbb{R}^{1,3} \oplus \Lambda^3 \mathbb{R}^{1,3} \oplus \Lambda^4 \mathbb{R}^{1,3}. 
\]
Given a basis $e_i$ for $\mathbb{R}^{1,3}$ and its dual $\mathcal{E}^i$, the Clifford algebra can thus be represented on $\Lambda^\bullet \mathbb{R}^{1,3} \ni \varpi$ by the wedge product operation and the contraction:
\begin{equation}
e_i \wedge \varpi + \iota_{\mathcal{E}^i} \varpi.
\label{Clifford-on-exterior-alg}
\end{equation}
Analogously we may take the vector space of sections $\Gamma(\Lambda^\bullet T^* \mathbb{R}^{1,3}) \cong \Gamma(\Lambda^\bullet T^*M)$ as our representation space. Then on a form $\omega^k$ of form degree $0 \leq k \leq 4$, the action \eqref{Clifford-on-exterior-alg} extends as the operation of wedging by $\mathrm{d}x^\mu$ plus contracting with $\partial_\mu$. Immediately, $q=\psi^\mu p_\mu$ is recognized to act on $\omega^k$ as:
\begin{equation}
	q \omega^k = -\mathrm{i}(\mathrm{d}x^\mu \wedge +  \, \eta^{\mu\nu}\iota_{\partial_\nu}) \partial_\mu \, \omega(x)_{\alpha_1 \dots \alpha_k} \mathrm{d}x^{\alpha_1} \wedge \dots \wedge \mathrm{d}x^{\alpha_k}= -\mathrm{i} (\mathrm{d} + \mathrm{d}^\dagger) \omega^k \, ,
\end{equation}
With $\mathrm{d}^\dagger = \ast \mathrm{d} \ast$ being the codifferential, and $\ast$ the Hodge star operator. Hence $q$ is exactly the "square root" of the Laplace-Beltrami operator $p^2  = - \square = - (\mathrm{d}\mathrm{d}^\dagger + \mathrm{d}^\dagger \mathrm{d})$ as it should be. Thus at each form degree, the conditions are:
\begin{equation}
	\begin{cases}
		\mathrm{d}^\dagger \omega^1 = 0 = \mathrm{d}^\dagger \omega^0 \\
		\mathrm{d} \omega^k + \mathrm{d}^\dagger \omega^{k+2} = 0 \quad \text{for $k=0,1,2$} \\
		\mathrm{d} \omega^3 = 0 = \mathrm{d} \omega^4
		\label{eq:n=1diff}
	\end{cases}
\end{equation}
while $\square \omega^k =0$ follows from the equations above. Remind that our metric is pseudoRiemannian. Contrary to the Riemannian case, there is no Hodge decomposition of a form into harmonic, exact and co-exact part in such setting. What holds instead, is a decomposition for a form $\omega^k$, with \emph{spacelike compact support}\footnote{A set $X$ is spacelike compact if it is contained in the causal influence set of a compact set Y.}, in case it is \emph{closed and co-closed}. Then $\omega^k = \mathrm{d}\omega^{k-1} + \mathrm{d}^\dagger \omega^{k+1}$, with $\omega^{k-1}, \omega^{k+1}$ spacelike compact forms satisfying $\mathrm{d} \omega^{k+1} = 0 $ and $\mathrm{d}^\dagger \omega^{k-1} =0$. The interested reader can consult \cite{Dappiaggi:2011zs}. However this is not an option here, because our differential equations \eqref{eq:n=1diff} are mixing the form degrees. An alternative representation on holomorphic forms, though in an Euclidean target space, is discussed in the Appendix.

\emph{Notice that even if in target space we set a flat Minkowski metric, all flat metrics (associated to a flat connection) are admissible. Indeed, the metric enters only in the definition of the codifferential. We will come back to this remark towards the end of \cref{sec:BRST-twists}.}

\section{Backgrounds}

The spinning particle model of the previous section describes a fermion or a set of forms after first quantization, and does not give any relevant information on the \emph{flat} backgrounds, which can be chosen to one's liking. On the contrary, a bosonic string fixes its target space to be the bosonic NS-NS sector of Supergravity, as a consequence of lack of conformal anomaly \cite{Callan:1985ia}. 
Furthermore, fully-fledged Supergravity is implied by consistency conditions for the quantization of a superstring \cite{Berkovits:2001ue}, in Berkovits' pure spinor formulation \cite{Berkovits:2000fe}. With similar methods applied to the RNS superstring \cite{Adamo:2014wea}, the bosonic NS-NS supergravity eom's were shown to follow. A common features of these two works was to deploy Becchi-Rouet-Stora-Tyutin quantization as their chosen quantization prescription. 

BRST is essentially cohomological, so a differential operator is present and can be easily twisted with suitable wanna-be background fields. Then nilpotency, even just on a restricted set of functions (for the classical case) or Hilbert space states (for the quantum case), must be checked. It yields conditions on the background fields, thus giving indications on which of these fields can couple to the particle/string and what equations of motion they must satisfy. 
Motivated by the results of \cite{Bonezzi:2020jjq} on NS-NS supergravity as an outcome of BRST of the $N=4$ spinning particle, \emph{our purpose is to investigate Ramond-Ramond fields in the background of a spinning particle}, which must be suitably rearranged for this endeavor. \textbf{Ramond-Ramond fields} (or fluxes) are basically a set of $n$-forms (with $n$ just even or just odd according to which string of type II is considered), related by Hodge duality and subjected to $\mathrm{d}$-closure. These two properties consequently imply that the forms must have zero divergence. Given the state of affairs explained in \cref{sec:Forms}, these conditions will not be hard to get from a spinning particle, given its deep ties with the D'Alembertian and the Dirac operator.

Let us first briefly review what BRST entails and then apply it to the spinning particle.

\subsection{BRST and twists by backgrounds} 
\label{sec:BRST-twists}

\textbf{Becchi-Rouet-Stora-Tyutin quantization} \cite{Becchi:1975nq} was developed in order to perform the path integral quantization of gauge theories, which are notoriously difficult to tackle with older methods because of the gauge fields (boundary states). To study functions on the singular symplectic quotient by the gauge symmetries, the idea is to introduce fictitious Grassmann coordinates so to form a \emph{resolution} of the quotient. These coordinates correspond to the symmetry generators in $\mathfrak{g}$ but with shifted parity (the \emph{antighosts}) as well as those of the dual $\mathfrak{g}^*$, again upon a shift in parity (the \emph{ghosts}). In this paper, the generators of $\mathfrak{g}$ are related to \eqref{symmetries}. The following step is to "embed" the resolution into a doubly graded chain complex: $\mathcal{C}=\sum_{p,q} \mathcal{C}^{p,q} = \sum_{p,q} C^\infty(M) \otimes \Lambda^p \mathfrak{g}^\ast[1] \otimes \Lambda^q \mathfrak{g}[1]$ (the number $1$ in square brackets refers precisely to the change in degree and thus parity). The differential operator acts by sending elements of fixed $p-q$ to $p-q+1$, hence it can increase the ghost number $p$ as well as decrease the antighost number $q$. The whole point of the construction is to achieve an isomorphism in cohomology: $H^\bullet(\mathcal{C}) \cong H^\bullet(\mathfrak{g}) \otimes C^\infty(M//G)$. At ghost degree $0$, one recovers the "physical" states while the gauge symmetries are placed at ghost degree $-1$.

In the present situation, to $H$ and $q$ given in \eqref{symmetries}, one assigns respectively $b, \beta \in \mathfrak{g}[1]$ and the only non-trivial bracket happens to be $[\beta, \beta] = b$. For the coalgebra then one has $\gamma, c$ such that $[\gamma, \beta] = 1 = \{b,c\} $. Our chosen \emph{polarization} is that $\beta$ and $b$ act with derivative action on a constant state, so that $\gamma$ and $c$ act like creation operators and generate the (ghost part of the) representation. Basically the representation of forms $\omega^n$ discussed in \cref{sec:Forms} is extended with $\mathbb{C}\llbracket \gamma, c\rrbracket$:
\begin{equation}
\sum_{k} \sum_{n=0}^{2} \gamma^k \omega_k^n + c \gamma^k \omega_k^n \, .
\label{states}
\end{equation}
The differential operator is a ghost degree 1 object, built with the constraints and a piece depending on the algebra to ensure nilpotency:
\begin{equation}
Q= - c \square + \gamma (\mathrm{d}+\mathrm{d}^\dagger) + \gamma^2 \partial_c \, . \label{BRST0-Q}
\end{equation}
In $\ker Q$ one gets
	\begin{align*}
	(\mathrm{d} + \mathrm{d}^\dagger)\omega^{\bullet}_{k} + \omega^{\bullet}_{k-1} = 0, \\
	-\square \omega^{\bullet}_k + (\mathrm{d} + \mathrm{d}^\dagger) \omega^{\bullet}_{k-1} = 0 
\end{align*}
The cohomology shows that for $k\geq 1$ one of the two complexes of forms can be eliminated in favor of the other. So this is "off-shell". It should be noticed, however, that at $k= 0$, which corresponds to ghost number 0, one finds $\mathrm{d} \omega^{n-1} + \mathrm{d}^\dagger \omega^{n+1} = 0$ as obtained with the quantization method in the previous section.

We already claimed, at the end of the previous section, that the backgrounds can support only flat connections. BRST, because of its cohomological nature, is a great playground to test this claim. A straightforward way to analyze backgrounds is by a twist of the de Rham differential in \eqref{BRST0-Q}. It should also be noticed that only the forms which appear in the Hilbert space can be used for the twist. Indeed the reason is that the BRST operator itself induces a homological vector field on the space of fields which can depend on the latter ones. 
A $1$-form is certainly present in our space of states \eqref{states}, so we can do the following: 
\[
 \mathrm{d} \rightsquigarrow \mathrm{d} + A \wedge =: \mathrm{d}_A, 
\]
and its adjoint $\mathrm{d}^\dagger_A$ is obtained by decorating it with the Hodge star on the left and the right. Recall that if $\varrho$ is a $p$-form, in $4$ dimensions $\ast \ast \varrho = - (-1)^{p(4-p)} \varrho$. In turn the differential must be, in first approximation:
\begin{align*}
Q_A =& \, c H_A + \gamma \left(\mathrm{d}_A + \mathrm{d}^\dagger_A \right) + \gamma^2 \partial_c \\
H_A :=  & \, -\square + \ast (\mathrm{d}^\dagger A) \wedge \ast + (\mathrm{d}^\dagger A)  \wedge + A \wedge \ast A \wedge \ast + \ast A \wedge \ast A \wedge \\
\, & \, +  \left([\mathrm{d}_A, \mathrm{d}_A ] \right) \wedge + \ast \left([\mathrm{d}_A, \mathrm{d}_A ] \right) \wedge \ast \, .
\end{align*}
Here $H_A$ is fixed by $\{\mathrm{d}_A + \mathrm{d}^\dagger_A, \mathrm{d}_A + \mathrm{d}^\dagger_A\} = H_A$. To ensure nilpotency, the quantum commutator of operators between $H_A$ and $\mathrm{d}_A + \mathrm{d}^\dagger_A$ must vanish.
Things will sensibly change depending on whether we consider an abelian or non-abelian theory. For the former, we obtain $\mathrm{d}^\dagger A =0$ as well as $\mathrm{d} A = 0$ (remarkably, $[\mathrm{d}_A,H_A]=0$ does not require further conditions so the $A^2$ term in $H_A$ does not have to be zero). Hence $A$ is de Rham closed and co-closed. 
 
Instead if we wanted to consider a non-abelian $1$-form, its field strength $ [\mathrm{d}_A, \mathrm{d}_A ]_+ = 2 \mathrm{d} A + [A,A] $ should again be set to zero, while $\mathrm{d}^\dagger_A A =0$ is just $ \mathrm{d}^\dagger A =0$ because $A^{\mu \, a} A^b_{\mu} f_{ab}{}^c = 0$, for $f_{ab}{}^c$ the structure constants.~In both cases, the conditions would be compatible with thinking of $A$ as a Ramond-Ramond 1-form field strength! However a complete match with the Ramond-Ramond fluxes requires also all the other admissible forms in higher degree. 
Unfortunately, in this setting it is not clear how to discuss deformations by a 2-form.
That is why, inspired by \cite{Sorokin:1988jor} and \cite{Sorokin:1988nj}, we resort to a twistorial description.

\subsection{Spin field and twistors}

Given that we are considering a supersymmetric worldline with a $4$-dimensional target space, the Spin group has two irreps distinguishable by their chirality\footnote{The chirality is the eigenvalue for the chirality operator: \[\frac{\text{id} +\psi_4}{2},\] where $\psi_4 := \psi^0\psi^1\psi^2\psi^3$. Since it is also a projector, its eigenvalues are only $\pm 1$.}. Hence one may consider introducing two pairs of \emph{conjugated} twistor-like variables, $(\theta^\alpha, \lambda_\beta)$ and $(\tilde\theta_{\dot\alpha}, \tilde\lambda^{\dot\beta})$ and could explore the consequences of assigning even parity to one pair, odd parity to the other: $[\theta^\alpha, \lambda_\beta] = \delta^\alpha_\beta$ and $\{\tilde\theta_{\dot\alpha}, \tilde\lambda^{\dot\beta}\} = \delta^{\dot\beta}_{\dot\alpha}$. An educated guess for the realization of the Gamma matrices could then be:
	\begin{equation}
	\psi^\mu := \theta^\alpha \sigma^\mu_{\alpha\dot\alpha} \tilde\lambda^{\dot\alpha} + \tilde\theta_{\dot\alpha} \tilde\sigma^{\mu\; \dot\alpha\alpha} \lambda_\alpha .
	\label{expr:Gamma-matrix}
\end{equation}
Now, if $\theta, \bar\theta$ create states out of a ground state, in the Fock space $F$ we would find bitwistors. These are nothing but forms in $\Omega^\bullet(M)$ written in twistorial notation. The action of $\psi^\mu$ extends from that on a single particle Hilbert space by Leibniz rule. However after some algebraic manipulations one would immediately recognize that on the Fock space,
\[ 
\{\psi^\mu, \psi^\nu \}
\]
equals $2\eta^{\mu\nu}$ only on the spinors, while on the bispinors/bitwistors equals $(2 + 2)\eta^{\mu\nu}$ because of Leibniz rule applied to the tensor representation. Abandoning reality and resorting to complex pairs $\sigma^a = \frac{1}{\sqrt{2}} (\sigma^0 + \mathrm{i}\sigma^1), \, \sigma^b := \frac{\mathrm{i}}{\sqrt{2}} (\sigma^2 + \sigma^3)$ and $\tilde\sigma^a = \frac{1}{\sqrt{2}} (\sigma^0 - \mathrm{i}\sigma^1), \, \tilde\sigma^b = \frac{\mathrm{i}}{\sqrt{2}} (\sigma^2 - \sigma^3)$ does not help either.

Following our work \cite{Boffo:2022pbs}, we will now enclose all the odd Grassmann parity into a $\mathbbm{Z}_2$ degree shifting operator $\uparrow$, so that now $\tilde\theta, \tilde\lambda$ are turned into Grassmann even spinors:
\[
[\lambda_\alpha, \theta^\beta] = \delta^\beta_\alpha, \quad [\tilde\lambda^{\dot\alpha}, \tilde\theta_{\dot\beta}] = \delta^{\dot\alpha}_{\dot\beta} \, ,
\]
The degree shifting operator, in combination with the twistor-like objects, can be thought as a \emph{spin field} for a spinning particle in the worldline. 
In string theory, a spin field is usually inserted at the endpoint of a branch cut to turn a Neveu-Schwarz state into a Ramond one. 
Since there are no branch cuts for functions on a line, as opposed to what can happen with complex maps from a conformal plane, this is the best that we can actually do to mimic this on the line. 

Furthermore, we choose $\lambda,\tilde\lambda$ to commute with $\uparrow$, while their conjugated variables are non-commuting with $\uparrow$. With appropriate insertions of $\uparrow$, our representation space $F^\prime$, limited to spinors and bispinors, contains specific polynomials in $\theta, \tilde\theta$:
\begin{align}
F^\prime = \left\langle v_\alpha \theta^\alpha, \quad \left(B_{\alpha\beta} +\varphi \epsilon_{\alpha\beta}\right) \ket{\alpha\beta}, \quad A_{\alpha}{}^{\dot\alpha} \ket{\alpha \dot\alpha} \right\rangle
\label{F'} \\
\ket{\alpha\beta} := \frac{1}{2} \left( \theta^\alpha (\uparrow \theta^\beta - \theta^\beta \uparrow) + \theta^\alpha(\theta^\beta - \uparrow \theta^\beta \uparrow)  \right) \, , \notag \\
\ket{\alpha \dot\alpha} := \frac{1}{2} \left( \theta^{\alpha} (\uparrow \tilde\theta^{\dot\alpha} -\tilde\theta^{\dot\alpha} \uparrow) + \theta^\alpha (\tilde\theta^{\dot\alpha} - \uparrow\tilde\theta^{\dot\alpha} \uparrow) \right) \, . \notag
\end{align}
With $\tilde{F}^\prime$ we will denote the chiral counterpart to $F^\prime$, where all the $\theta$'s are traded for $\tilde\theta$ and vice versa. With these choices, the anticommutator of the Gamma matrices \eqref{expr:Gamma-matrix} $\psi^\mu \uparrow$ yields now:
\begin{equation}
	\{\psi^\mu \uparrow, \psi^\nu \uparrow\} = 2\eta^{\mu\nu} \left(\theta^\alpha \lambda_\alpha + \tilde\theta_{\dot\alpha} \tilde\lambda^{\dot\alpha}  \right) + f_1(\tilde\theta) \lambda \lambda + f_2(\theta) \tilde\lambda \tilde\lambda + f_3(\theta,\tilde\theta) \tilde\lambda \lambda \, ,
	\label{Clifford+junk}
\end{equation}
but one can be easily convinced that the last three terms in \eqref{Clifford+junk} drop after evaluating the above expression on any state of $F^\prime$ and $\tilde{F}^\prime$. Then $(\theta^\alpha \lambda_\alpha + \tilde\theta_{\dot\alpha} \tilde\lambda^{\dot\alpha})$ just fixes the representation space to be invariant under rotations in the space of spinors with the same chirality. In the end, one has managed to retrieve: 
\[
\{\psi^\mu \uparrow, \psi^\nu \uparrow\}\vert_{F^\prime, \tilde{F}^\prime} = 2 \eta^{\mu\nu} \, .
\]

As customary in BRST, we should now extend these Weyl and anti-Weyl spinors and bispinors, 
to a representation of the ghost algebra, however for our current investigations we can just focus on the ghost degree zero states. These are in $\ker Q$ if
\begin{align}
	Q F^\prime = \gamma \psi^\mu (-\mathrm{i}\partial_\mu) F^\prime = 0 \\
	\begin{cases}
		\slashed{\partial} \nu = 0\\
		\mathrm{d} A = 0 = \mathrm{d}^\dagger A, \\
		\mathrm{i}\mathrm{d}^\dagger B + \mathrm{d}\varphi - \frac{1}{2}\ast\mathrm{d} B = 0 .
	\end{cases}\label{coho}
\end{align}
Hence we recover once more the Weyl equation for our Weyl spinor $\nu$.
Regarding the remaining equations in \eqref{coho}, they follow from the \emph{Fierz identities} with the Pauli matrices\footnote{Please refer to the appendix for a collection of useful identities.}, and naturally separate according to the form degree, but also into imaginary and real part.~Hence at ghost degree zero we find odd degree forms which are closed and co-closed, $\mathrm{d}A =0 = \mathrm{d}^\dagger A$. Then if one assumes that $B$ has real values, we can require it to be divergence-free. To solve for the real part of the last equation in \eqref{coho} we can finally use a duality condition. Indeed for our present discussion it is rather crucial to half the number of degrees of freedom, by making $p$-forms Hodge dual to $(4-p)$-forms (or self-dual in the case of the 2-form). Duality and either closure or co-closure imply the remaining differential equation: for instance, taking for concreteness $A^{(1)} = \ast A^{(3)}$ and $\mathrm{d}A^{(1)} = 0 = \mathrm{d}A^{(3)}$, then it is guaranteed that they have zero divergence. Coming back to the real part of the bottom equation in \eqref{coho}, if we impose self-duality in spacetime, $B = \ast B$, then $\mathrm{d}\varphi =0$ separately. 
Hence $\varphi, A \equiv A^{(1)}$ and $B$ can be interpreted as Ramond-Ramond field strengths. Certainly a field strength of form degree $0$ (like $\varphi$) is not meaningful in the realm of de Rham cohomology (it would require K-theory, but this is beyond the scopes of this note). As done in string theory and supergravity, we will avoid any complication caused by $\varphi$ by taking it to be constant.

At higher ghost numbers the theory is still off-shell as before. Notably the equations \eqref{coho} do not require the Hamiltonian constraint (or zero mass constraint), so we can actually drop it altogether, and consider only the chiral supercharge
\begin{equation}
	\mathbf{q} = \tilde\theta_{\dot \alpha} \tilde\sigma^{\mu \; \dot \alpha \beta} \lambda_\beta \, p_\mu  \uparrow .
	\label{chiral_supercharge}
\end{equation}
Basically $H^0_Q(M,\mathbb{R}) = H^0_{\mathbf{q}}(M, \mathbb{R})$. Twistings of this operator are easy to handle. 

\subsection{R-R backgrounds}

We are now all set and can embark in the study of deformations by background fields. The focus will be on the chiral supercharge $\mathbf{q} = \tilde\theta \slashed{\tilde{p}} \lambda \uparrow$. Such operator has an action on $F^\prime$ \eqref{F'} while $\tilde F^\prime$ is annihilated by $\mathbf{q}$. The deformations that we are keen on studying are:
\[
\delta \mathbf{q}_{\tilde B} := \tilde \theta_{\dot\alpha} \tilde B^{\dot\alpha\dot\beta} \tilde\lambda_{\dot\beta} \uparrow, \qquad \delta \mathbf{q}_A := \tilde\theta_{\dot\alpha} \tilde{A}^{\dot\alpha}{}_\beta \lambda^\beta \uparrow \, ,
\]
with $\tilde B$ and $\tilde A$ being a priori just a $2$-form and a $1$-form written in twistorial notation. The possibility of studying "covariant Dirac operators" with forms in different form degree is a remarkable feature of the twistorial description.

Now,
\[
\{ \mathbf{q} + \delta\mathbf{q}_A, \mathbf{q} + \delta\mathbf{q}_A \} = 2 \{\mathbf{q},\delta\mathbf{q}_A\} + \{\delta \mathbf{q}_A, \delta \mathbf{q}_A \}
\]
is going to be automatically nilpotent on every state in $F^\prime$ and its chiral counterpart. This a simple consequence of having two annihilators ($\lambda_\alpha \lambda_\beta = \partial_{\theta^\alpha}\partial_{\theta^\beta}$) on the right, and it is a radical departure from the behavior observed so far. Shall a larger space of states not be annihilated by them, the condition for nilpotency is just $\mathrm{d}A=0$.

Concerning $\delta\mathbf{q}_{\tilde B}$ things get quite intriguing. While the linear deformation $\{\delta\mathbf{q}_{\tilde B}, \mathbf{q}\}$ is zero on the locus of the equations of motion for the fields in $F^\prime$ and $ \tilde F^\prime$, nilpotency cannot be achieved because:
\begin{equation}
	 \{\delta\mathbf{q}_{\tilde B}, \delta\mathbf{q}_{\tilde B}\} \propto \tilde\theta \left( \ast \tilde B \wedge \ast \tilde B + \tilde B \wedge \tilde B + \tilde B \circ g^{-1} \tilde B \right) \tilde{\lambda}
	 \label{B-defo}
\end{equation}
as seen by thoroughly using \eqref{square}. Since it depends linearly on $\tilde\lambda$, \eqref{B-defo} has still a non-zero action on $\tilde F^\prime$. We cannot ignore $F^\prime$ and simply project our theory into the chiral half $F^\prime$ because in that case it will not contain $\tilde B$. Hence we would be forced to set $\tilde B=0$. Nevertheless it is interesting to treat this as a small deformation. Expanding the field $A$ of the Hilbert space in a small parameter $A = A_{(0)} + s A_{(1)} + \dots$, in the kernel of $\mathbf{q} + \delta\mathbf{q}_B$ at first order in $s$ we find:
\begin{align}
	\mathrm{d}^\dagger A_{(1)} =& - \ast \left(\tilde{B} \wedge \ast \tilde{B}_{(0)}\right) \\
	\mathrm{d} A_{(1)} =& \, \tilde{B} \circ g^{-1} \circ \tilde{B}_{(0)}
\end{align}
We deem these equation to stem from a BF-type \cite{Dijkgraaf:1989pz} or Chern-Simons-type theory. 

\section{Conclusions and outlook}

In this short note, after reviewing the $N=1$ spinning particle and especially its quantization into spinors and forms, we studied Ramond-Ramond fields in the background. The construction relies on twistor-like objects and on BRST quantization. In the BRST operator, the supercharge (associated to supersymmetry invariance) can be twisted by $1$- and $2$-forms in the twistorial notation. Nilpotency of the newly defined BRST operator does not have to hold tout-court, but can be verified on the states in the Fock space of the quantized spinning particle. Thanks to a spin field for the worldline, we could show that a \emph{dynamical} $1$-form is admitted: there are no obstructions for a covariant derivative constructed with it. The equations of motion for the $1$-form are an independent piece of information which does not follow from nilpotency. Instead a deformation by the $2$-form did not lead to a nilpotent operator on the Fock space. Nevertheless, when this finite deformation is turned into an infinitesimal one we obtain some intriguing differential equations.

In light of the analysis presented here, which partly summarizes \cite{Boffo:2022pbs}, the $N=1$ spinning particle in the twistorial description does yield some sensible results about Ramond-Ramond fluxes in the background. However we believe that spinning particles with enhanced supersymmetry should improve these outcomes and lead to the full set of R-R fields and their equations. Considering models with explicit supersymmetry in the target space e.g.~the \emph{superembedding model} \cite{Sorokin:1999jx}, should noticeably help to get the fermionic fields too.

\section*{A Fierz identities}

For $\eta^{\mu\nu} = \text{diag}(1, -1, -1 , -1)$,
\begin{align}
	\left(\sigma^{\mu\nu}\right)_{\alpha}{}^\beta &= \frac{\mathrm{i}}{4} \left(\sigma^\mu_{\alpha\dot\gamma} \tilde\sigma^{\nu \, \dot\gamma \beta} - \sigma^\nu_{\alpha\dot\gamma} \tilde\sigma^{\mu \; \dot\gamma\beta}\right) \\
	\tilde{\sigma}^{\rho \, \dot\alpha \alpha} \sigma^{\mu \nu}_{\alpha \beta} &= \begin{cases}
		\frac{\mathrm{i}}{2} (\eta^{\rho \nu} \tilde{\sigma}^{\mu} - \eta^{\rho \mu}\tilde{\sigma}^{\nu})^{\dot\alpha}{}_\beta - \frac{1}{2} \epsilon^{\rho \mu \nu \delta}\tilde{\sigma}_{\delta}^{\dot \alpha \gamma} \epsilon_{\gamma \beta} \; \; \; \text{if} \, \rho = j
		\label{tripl} \\
		\frac{\mathrm{i}}{2} (\eta^{\rho \nu} \tilde{\sigma}^{\mu} + \eta^{\rho \mu}\tilde{\sigma}^{\nu})^{\dot\alpha}{}_\beta - \frac{1}{2} \epsilon^{\rho \mu \nu \delta}\tilde{\sigma}_{\delta}^{\dot \alpha \gamma} \epsilon_{\gamma \beta} \; \; \; \text{if} \, \rho=0
	\end{cases}\\
	\delta_{\dot\beta}^{\dot\alpha}(\sigma^\mu)_{\alpha\dot\alpha}(\tilde\sigma^\nu)^{\dot\beta\beta} &= \eta^{\mu\nu}\delta^{\beta}_{\alpha}-2\mathrm{i}[\sigma^{\mu\nu}]^{\beta}{}_{\alpha}\\
	(\tilde{\sigma}^{\mu \nu} \tilde{\sigma}^{\rho \sigma})^{\dot\alpha \dot \beta} &= \frac{1}{4} \left(\eta^{\mu \sigma} \eta^{\rho\nu} -\eta^{\mu\rho}\eta^{\nu\sigma}\right) \epsilon^{\dot \alpha \dot\beta} + \frac{\mathrm{i}}{4} \epsilon^{\mu\nu\rho\sigma} \epsilon^{\dot \alpha \dot\beta} \\ 
	\, & \; \,  + \frac{\mathrm{i}}{2} \left(\eta^{\sigma\nu}\tilde{\sigma}^{\mu\rho} - \eta^{\nu\rho} \tilde{\sigma}^{\mu\sigma} +\eta^{\mu\rho}\tilde{\sigma}^{\nu\sigma} - \eta^{\mu\sigma}\tilde{\sigma}^{\nu\rho}\right)^{\dot\alpha \dot\beta}, \label{square}
\end{align}

\section*{B Alternative representation on holomorphic forms}

To build such representation, let us first swap the pseudoRiemannian target space for a Riemannian one, with Euclidean metric. Many complications are avoided when the metric is positive definite. Then one can construct two complex conjugated pairs of $\psi$'s: 
\begin{equation}
	\Psi^p = \frac{1}{\sqrt{2}}   \left(\psi^{2p} + \mathrm{i} \psi^{2p+1}\right) , \quad  \bar\Psi^p = \frac{1}{\sqrt{2}}  \left(\psi^{2p} - \mathrm{i} \psi^{2p+1} \right),  \quad p=0,1 \label{complex-gamma}
\end{equation}
This linear redefinition is a homomorphism of the Clifford algebra, yielding the new brackets:
\begin{align}
	\, & \{\Psi^k, \bar\Psi^j\} = 2\delta^{kj}, \quad \{\Psi^i,\Psi^k \} = 0 = \{\bar\Psi^j, \bar\Psi^i\} \, . \label{fermionic-oscillator-alg}
\end{align} 
Considering \eqref{complex-gamma} themselves as a representation, $\Psi^0, \Psi^1$ transform in the fundamental of $\mathfrak{su}(2)$, while $\bar\Psi^0, \bar\Psi^1$ in the antifundamental. Then a module for $\mathfrak{so}(4)$ inside $\text{Spin}(4)$, constructed from the lowest weight state (or ground state in physics), consists of $\mathfrak{su}(2)$ totally antisymmetric tensors due to \eqref{fermionic-oscillator-alg}. Thus the states in the Hilbert space are: 
\begin{equation*}
	\Phi(x) \ket{0} 
	, \quad C_k (x) \Psi^k \ket{0} 
	, \quad B_{01}(x) (\Psi^0\Psi^1 - \Psi^1\Psi^0) \ket{0} . 
\end{equation*}
So $\Phi$ is the ground state singlet, $C_k (x) \Psi^k$ is the $2$ (vectorial representation), and the last state is the singlet in $2\otimes 2 = 3 \oplus 1$. One should then impose the dynamics, determined by the constraints $p^2=0$ and $\psi^\mu p_\mu =0 $. For this scope, a convenient realization of the algebra \eqref{fermionic-oscillator-alg} is by the following assignation:
\begin{equation}
\Psi^i \cong \mathrm{d}z^i \wedge , \quad \bar\Psi^i \cong h^{\bar{i} k}\iota_{\partial_{z^k}} \, .
\end{equation}
Indeed the exterior product with a holomorphic differential and the contraction by a holomorphic vector field, together with the hermitian metric $h^{\bar{i}k} = (h^{-1})^{\bar{i}k} = \delta^{\bar{i} k}$, replicate \eqref{fermionic-oscillator-alg}. We are thus locally seeing the 4-dimensional manifold $\mathbb{R}^{4}$ as a 2-dimensional almost complex manifold $U$. Upon this identification, the momenta in the cotangent space must also be arranged in complex pairs:
\[
p_\mu \rightarrow (p_{z^0}, p_{z^1}) = \left(\frac{1}{\sqrt{2}}(p_0 - \mathrm{i} p_1), \frac{1}{\sqrt{2}} (p_2 - \mathrm{i} p_3)\right)
\]
In turn, the Dirac operator $\psi^\mu p_\mu$, expressed in terms of $\Psi, \bar\Psi$ and the complex momenta, and letting $p_z \rightarrow -\mathrm{i} \partial_z$, is now explicitly realized as:
\begin{equation}
\psi^\mu p_\mu \mapsto - \mathrm{i} \left(\mathrm{d}z^k \partial_{z^k} + h^{l\bar{k}}\iota_{\partial_{z^l}} \partial_{\bar{z}^{\bar{k}}}\right) \equiv - \mathrm{i} \left(\partial + \partial^\dagger\right), 
\end{equation}
with Dolbeault differential and codifferential. Instead the Laplacian is now given by the holomorphic part of the complex Laplacian:
\[
p^2 \mapsto - \Delta_\partial
\]
On the states, the Dirac operator constraint tells us:
\begin{equation}
\partial C(z +\bar z) =0, \qquad \partial^\dagger C(z +  \bar z) = 0, \qquad \partial^\dagger B(z + \bar z) + \partial \Phi(z + \bar z) = 0, \label{clo-coclo}
\end{equation}
which the reader can immediately recognize to be already in the kernel of the holomorphic Laplacian. Hence the 1-form is $\partial$-closed and divergence-less, while the remaining equation relates the holomorphic divergence of the 2-form to the Dolbeault differential of the function. 

\section*{Acknowledgments}
I would like to thank the organizers of the workshop "Noncommutative and generalized geometry in string theory, gauge theory and related physical models" for the opportunity to present my results and for a successful workshop. A huge thank you to the local organizers for the relaxed atmosphere and for putting up an incredible program rich in cultural and sport activities. \\
\noindent I have benefited from discussions with Mauro Mantegazza, Svatopluk Krysl and Ondra Hul\'{i}k, to whom I am very grateful. I am thankful to Ivo Sachs for working together on the paper which prompted this short article. Financial support from GA\v{C}R grant EXPRO 19-28268X and from a Riemann Fellowship is acknowledged.

\end{document}